\documentclass[12pt]{article}
\newcommand{\ba}{\begin{array}}
\newcommand{\ea}{\end{array}}
\newcommand{\nn}{\nonumber\\}

\newcommand{\res}{\mbox{res}}

\newcommand{\del}{\partial}

\newcommand{\rar}{\rightarrow}
\newcommand{\lar}{\leftarrow}

\newcommand{\fr}{\frac}

\newcommand{\scr}{\scriptsize}

\newcommand{\st}{\stackrel}
\makeatletter
\@addtoreset{equation}{section}

\makeatother

\begin{document}

\begin{titlepage}
\null
\begin{flushright}
UT-Komaba/03-18
\\
hep-th/0311206
\end{flushright}

\vskip 1.5cm
\begin{center}

 {\Large \bf Commuting Flows and Conservation Laws}

\lineskip .75em
 
 {\Large \bf for Noncommutative Lax Hierarchies}

\vskip 1.7cm
\normalsize

 {\large Masashi Hamanaka}

\vskip 1.5cm

        {\it Institute of Physics, 
         University of Tokyo, Komaba,\\
         Meguro-ku, Tokyo 153-8902, Japan\footnote{Present address:
Graduate School of Mathematics, Nagoya University,
                     Chikusa-ku, Nagoya, 464-8602, JAPAN
(E-mail: hamanaka@math.nagoya-u.ac.jp)}}


\vskip 1.5cm

{\bf \large Abstract}

\end{center}

We discuss commuting flows and conservation laws
for Lax hierarchies on noncommutative spaces
in the framework of the Sato theory.
On commutative spaces, the Sato theory has revealed
essential aspects of the integrability
for wide class of soliton equations
which are derived from the Lax hierarchies 
in terms of pseudo-differential operators.
Noncommutative extension of the Sato theory
has been already studied by the author and Kouichi Toda,
and the existence of various noncommutative Lax 
hierarchies are guaranteed.
In this paper, we present 
conservation laws for the noncommutative Lax hierarchies
with both space-space and space-time noncommutativities
and prove the existence of 
infinite number of conserved densities.
We also give the explicit representations of them 
in terms of Lax operators.
Our results include noncommutative versions of
KP, KdV, Boussinesq, coupled KdV, Sawada-Kotera, 
modified KdV equations and so on.

\end{titlepage}
\clearpage
\baselineskip 6.8mm

\section{Introduction}

Non-Commutative (NC) extension of
field theories has been studied intensively 
for the last several years \cite{NC}.
NC gauge theories are equivalent to
ordinary gauge theories
in the presence of background magnetic fields
and succeeded in revealing various aspects 
of them \cite{YM}.
NC solitons especially play important roles 
in the study of D-brane dynamics, 
such as the confirmation of Sen's conjecture 
on tachyon condensation \cite{Harvey}.
One of the distinguished features of NC theories
is resolution of singularities.
This gives rise to various new physical objects
such as $U(1)$ instantons and 
makes it possible to analyze 
singular configurations as usual.

NC extension of integrable equations such as 
the Korteweg-de Vries (KdV) equation \cite{KdV}
is also one of the hot topics \cite{CFZ}-\cite{GMPT}.
These equations imply no gauge field and
NC extension of them perhaps might have  
no physical picture or no good property on integrabilities.
To make matters worse,
the NC extension of $(1+1)$-dimensional equations
introduces infinite number of time derivatives,
which makes it hard to discuss or define the integrability.
However, some of them actually possess integrable properties, 
such as the existence of infinite number 
of conserved quantities \cite{DiMH2, DiMH3,DiMH4,GrPe}
and the linearizability \cite{HaTo2, MaPa}
which are widely accepted as definition of 
completely integrability of equations.
Furthermore, a few of them can be derived from 
NC (anti-)self-dual Yang-Mills (YM) equations
by suitable reductions \cite{Legare, HaTo2, NiRa}.
This fact may give some physical meanings and good properties
to the lower-dimensional NC field equations
and makes us expect that 
the Ward conjecture \cite{Ward} still holds 
on NC spaces \cite{HaTo}.
So far, however, those equations have been
examined one by one. Now it is very natural to discuss
their integrabilities in more general framework.

The author and Kouichi Toda have studied
systematic NC extension of integrable systems 
\cite{HaTo, HaTo2, HaTo3}.
In the previous paper \cite{HaTo3},
we have obtained wide class of NC Lax hierarchies 
which include various NC versions of soliton equations
in the framework of the Sato theory \cite{Sato}.
On commutative spaces, the Sato theory 
is known to be one of the most beautiful
theories of solitons and reveals 
essential aspects of the integrability,
such as, the construction of exact multi-soliton solutions,
the structure of the solution space,
the existence of infinite conserved quantities,
and the hidden symmetry of them.
In the Sato theory, the soliton equations are described by
Lax hierarchies in terms of pseudo-differential operators.

In the present paper,
we prove the existence of infinite conserved quantities
for Lax hierarchies on NC spaces
in the framework of the Sato theory.
We show 
the conservation laws
for them and give the explicit representations
with both space-space and space-time noncommutativities.
This suggests that the NC soliton equations 
are also completely integrable
and infinite-dimensional symmetries 
would be hidden.
Our results include wide class of NC soliton equations,
such as, NC versions of
Kadomtsev-Petviashvili (KP) \cite{KaPe}, KdV, 
Boussinesq \cite{Boussinesq}, coupled KdV \cite{HiSa}, 
Sawada-Kotera \cite{SaKo}, modified KdV (mKdV) equations 
and so on.

\section{Comments on Noncommutative Field Theories}

NC spaces are defined by 
noncommutativity of the coordinates:
\begin{eqnarray}
\label{nc_coord}
[x^i,x^j]=i\theta^{ij},
\end{eqnarray}
where $\theta^{ij}$ are real constants and 
called the {\it NC parameters}.

NC field theories are obtained from given commutative
field theories by exchange of
ordinary products in the commutative field theories 
for {\it star-products}.
The star-product is defined 
for ordinary fields on commutative spaces.
On Euclidean spaces, it is explicitly given by 
\begin{eqnarray}
f(x)\star g(x)&:=&f(x)
\exp{\left(\frac{i}{2}\st{\lar}{\partial}_i\theta^{ij}
\st{\rar}{\partial}_j\right)}g(x)\nonumber\\
&=&f(x)g(x)+\frac{i}{2}\theta^{ij}\partial_if(x)\partial_jg(x)
+{\cal O}(\theta^2),
\label{star}
\end{eqnarray}
where $\del_i:=\del/\del x^{i}$.
This explicit representation is known 
as the {\it Moyal product} \cite{Moyal}.

The star-product possesses 
associativity: $f\star(g\star h)=(f\star g)\star h$,
and returns back to the ordinary product 
in the commutative limit: $\theta^{ij}\rar 0$.
The modification of the product  makes the ordinary 
spatial coordinates ``noncommutative,'' 
that is, $[x^i,x^j]_\star:=x^i\star x^j-x^j\star x^i=i\theta^{ij}$.

We note that the fields themselves take c-numbers values
and the differentiation and the integration for them 
are well-defined as usual.
NC field theories should be interpreted as
deformed theories from commutative ones.
One of nontrivial points in the NC extension
is order of nonlinear terms. 
The difference between commutative equations
and the NC equations arise as commutators of fields
which sometimes become serious obstructions.

\vspace{3mm}

Here we point out a special property of the NC commutators
of fields. It is convenient to introduce the following symbol
\begin{eqnarray}
 P:=\frac{1}{2}\st{\lar}{\partial}_i\theta^{ij}
\st{\rar}{\partial}_j,
\end{eqnarray}
and the {\it Strachan product} \cite{Strachan}
\begin{eqnarray}
 f(x)\diamond g(x):=f(x) \left(\sum_{s=0}^{\infty}
\fr{(-1)^s}{(2s+1)!}P^{2s}\right)g(x).
\end{eqnarray}

A commutator of fields is straightforwardly calculated as follows
\begin{eqnarray}
 [f(x),g(x)]_\star&=&f(x)(e^{iP}-e^{-iP}) g(x)=:2i f(x)(\sin P) g(x)\nn
&=&-\theta^{ij}\partial_i f(x)\diamond \partial_j g(x)\nn
&=&-\theta^{ij}\partial_i (f(x)\diamond \partial_j g(x)).
\end{eqnarray}
In the second line, we use the fact 
that $\sin P$ is the composite of $P$ and ``$P^{-1}\sin P$''
and the Strachan product ``$\diamond$'' corresponds to the latter.
This derivation was first presented
by Dimakis and M\"uller-Hoissen
in order to generate infinite number of 
conserved densities of
the NC non-linear Schr\"odinger (NLS) equation \cite{DiMH2},
the NC KdV equation \cite{DiMH3} and 
the NC extended matrix-NLS equation \cite{DiMH4}.
Here more widely, we would like to stress that
{\it commutators of fields on NC spaces always appear 
as total derivatives in the NC directions}.
This will be crucial in derivation of conservation laws in Sec. 5.

As a consequence, we can prove
\begin{eqnarray}
\label{integral}
 \int d^{D}x~ f(x)\star g(x)=\int d^{D}x~ f(x)g(x),
\end{eqnarray}
where the integration is taken 
in all NC directions.

\section{Noncommutative Lax Hierarchies in Sato's Framework}

In this section, we derive various
NC Lax equations
in terms of pseudo-differential operators
which include negative powers 
of differential operators.
We note that the present discussion in this section 
can be applied to more general cases where the
products are not necessarily the star-products 
but noncommutative associative products with differentiations,
which has been already discussed in e.g. \cite{Kupershmidt}.
However, we believe that some explicit examples here 
are new equations and would be useful for further studies.

An $N$-th order (monic) pseudo-differential operator $A$
is represented as follows
\begin{eqnarray}
 A=\del_x^N + a_{N-1}\del_x^{N-1}+ \cdots 
+ a_0 +a_{-1}\del_x^{-1}+a_{-2}\del_x^{-2}+\cdots.
\end{eqnarray}
Here we introduce useful symbols:
\begin{eqnarray}
 A_{\geq r}&:=& \del_x^N + a_{N-1}\del_x^{N-1}+ \cdots + a_{r}\del_x^{r},\\
 A_{\leq r}&:=& A - A_{\geq r+1} 
 = a_{r}\del_x^{r} + a_{r-1}\del_x^{r-1} +\cdots,\\
 \res_{r} A &:=& a_{r}.
\end{eqnarray}
The symbol $\res_{-1} A$ is especially called the {\it residue} of $A$.

The action of a differential operator $\partial_x^n$ on
a multiplicity operator $f$ is formally defined 
as the following generalized Leibniz rule:
\begin{eqnarray}
 \partial_x^{n}\cdot f:=\sum_{i\geq 0}
\left(\begin{array}{c}n\\i\end{array}\right)
(\partial_x^i f)\partial^{n-i},
\end{eqnarray}
where the binomial coefficient is given by
\begin{eqnarray}
\label{binomial}
 \left(\begin{array}{c}n\\i\end{array}\right):=
\frac{n(n-1)\cdots (n-i+1)}{i(i-1)\cdots 1}.
\end{eqnarray}
We note that the definition of the binomial coefficient (\ref{binomial})
is applicable to the case for negative $n$,
which just define the action of 
negative power of differential operators.
The examples are, 
\begin{eqnarray}
 \partial_x^{-1}\cdot f&=& 
f\partial_x^{-1}-f^\prime\partial_x^{-2}
+f^{\prime\prime}\partial_x^{-3}-\cdots,\nn
 \partial_x^{-2}\cdot f&=& 
f\partial_x^{-2}-2f^\prime\partial_x^{-3}
+3f^{\prime\prime}\partial_x^{-4}-\cdots,\nn
 \partial_x^{-3}\cdot f&=& 
f\partial_x^{-3}-3f^\prime\partial_x^{-4}
+6f^{\prime\prime}\partial_x^{-5}-\cdots,
\end{eqnarray}
where $f^\prime:=\del f/\del x, 
f^{\prime\prime}:=\del^2 f/\del x^2$ and so on,
and $\partial_x^{-1}$ in the RHS
acts as an integration operator $\int^x dx$.

The composition of pseudo-differential operators
is also well-defined and the total set 
of pseudo-differential operators forms 
an operator algebra.
For more on pseudo-differential operators 
and the Sato theory, see e.g. \cite{DJM}-\cite{OSTT}.

\vspace{3mm}

Let us introduce a Lax operator
as the following first-order pseudo-differential operator:
\begin{eqnarray}
 L = \partial_x + u_1 + u_2 \partial_x^{-1} 
 + u_3 \partial_x^{-2} + u_4 \partial_x^{-3} + \cdots,
\end{eqnarray}
where the coefficients $u_k$ ($k=1,2,\ldots$) are functions 
of infinite variables $(x^1,x^2,\ldots)$ with $x^1\equiv x$:
\begin{eqnarray}
 u_k=u_k(x^1,x^2,\ldots).
\end{eqnarray}
The noncommutativity is arbitrarily introduced for
the variables $(x^1,x^2,\ldots)$ as Eq. (\ref{nc_coord}) here.

The Lax hierarchy is defined in Sato's framework as
\begin{eqnarray}
 \del_m L = \left[B_m, L\right]_\star,~~~m=1,2,\ldots,
\label{lax_sato}
\end{eqnarray}
where the action of $\del_m$ on the pseudo-differential operator $L$
should be interpreted to be coefficient-wise, 
that is, $\del_m L :=[\del_m,L]$ or $\del_m \del_x^k=0$.
The operator $B_m$ is given by
\begin{eqnarray}
 B_m
:=(\underbrace{L\star \cdots \star L}_{
m{\scriptsize\mbox{ times}}})_{\geq r}=:(L^m)_{\geq r},
\end{eqnarray}
where $r$ is 0 for $u_1=0$ and 1 for $u_1\neq 0$
as commutative cases \cite{GeDi, Kupershmidt2}.
The Lax hierarchy gives rise to a set of 
infinite differential equations 
with respect to infinite kind of fields
from the coefficients
in Eq. (\ref{lax_sato}) for a fixed $m$.
Hence it contains huge amount of 
differential (evolution) equations for all $m$.
The LHS of Eq. (\ref{lax_sato}) becomes $\del_m u_k$
which shows a flow in the $x^m$ direction. 

If we put the constraint $L^l=B_l$
on the Lax hierarchy (\ref{lax_sato}),
we get infinite set of NC (reduced) Lax hierarchies.
We can easily show
\begin{eqnarray}
\label{Nl}
\frac{\partial u_k}{\partial x^{Nl}}=0,
\end{eqnarray}
for all $N,k$ because 
\begin{eqnarray}
 \fr{dL^l}{dx^{Nl}}=[B_{Nl},L^l]_\star=[(L^{l})^N,L^l]_\star=0,
\end{eqnarray}
which implies Eq. (\ref{Nl}).
The reduced NC hierarchy is called the {\it l-reduction}
of the NC KP hierarchy. 
This time, the constraint $L^l=B_l$ gives simple relationships
which make it possible to represent 
infinite kind of fields $u_{l-r+1},u_{l-r+2},u_{l-r+3},\ldots$
in terms of $(l-1)$ kind of fields 
$u_{2-r},u_{3-r},\ldots, u_{l-r}$. (cf. Appendix A.)

{}From now on, let us see that those equations 
in the Lax hierarchy contain various soliton equations 
with some constraints.
We discuss it separately in the following two cases:
$u_1=0$ ($r=0$) case and $u_1\neq 0$ ($r=1$) case.
Some of them are already discussed in \cite{HaTo3}.
For commutative discussions, see also \cite{MSS}.

\vspace{2mm}
\noindent
\underline{For $u_1=0$ ($r=0$)}
\noindent
\vspace{3mm}

In this case, the Lax hierarchy (\ref{lax_sato}) 
is just the {\it NC KP hierarchy} which 
includes the NC KP equation \cite{Paniak, Kupershmidt}.
Let us see it explicitly.

\begin{itemize}

\item NC KP hierarchy

The coefficients of each powers of (pseudo-)differential
operators in the Lax hierarchy (\ref{lax_sato}) yield a series 
of infinite NC ``evolution equations,'' that is, for $m=1$
\begin{eqnarray}
\partial_x^{1-k})~~~ \del _1 u_{k}=u_{k}^\prime,~~~k=2,3,\ldots 
~~~\Rightarrow~~~x^1\equiv x,
\end{eqnarray}
for $m=2$
\begin{eqnarray}
\label{KP_hie}
\partial_x^{-1})~~~\del_2 u_{2}
&=&u_2^{\prime\prime}+2u_{3}^{\prime},\nonumber \\
\partial_x^{-2})~~~
\del_2 u_{3}&=&u_3^{\prime\prime}+2u_4^{\prime}
+2u_2\star u_2^\prime +2[u_2,u_3]_\star,\nonumber \\
\partial_x^{-3})~~~
\del_2 u_{4}&=&u_{4}^{\prime\prime}+2u_{5}^{\prime}
+4u_3\star u_2^\prime-2u_2\star u_2^{\prime\prime}
+2[u_2,u_4]_\star,\nn
\partial_x^{-4})~~~\del_2 u_{5}&=&\cdots,
\end{eqnarray}
and for $m=3$
\begin{eqnarray}
\label{3flow}
\partial_x^{-1})~~~
\del_3 u_{2}&=&u_{2}^{\prime\prime\prime}+3u_3^{\prime\prime}
+3u_4^{\prime}+3u_2^\prime\star u_2+3u_2\star u_2^\prime,
\nonumber\\
\partial_x^{-2})~~~
\del_3 u_{3}&=&u_{3}^{\prime\prime\prime}+3u_{4}^{\prime\prime}
+3u_{5}^\prime+6u_{2}\star u_{3}^\prime+3u_2^\prime\star u_3 
+3u_3\star u_2^\prime+3[u_2, u_4]_\star,\nn
\partial_x^{-3})~~~
\del_3 u_{4}&=&u_{4}^{\prime\prime\prime}+3u_{5}^{\prime\prime}
+3u_{6}^\prime+3u_{2}^\prime \star u_{4}+3u_2\star u_4^\prime
+6u_4\star u_2^\prime\nn
&&-3u_2\star u_3^{\prime\prime}
-3u_3\star u_2^{\prime\prime}+6u_3\star u_3^{\prime}
+3[u_2,u_5]_\star+3[u_3,u_4]_\star,\nn
\partial_x^{-4})~~~\del_3 u_{5}&=&\cdots.
\end{eqnarray}
These just imply the $(2+1)$-dimensional 
NC KP equation \cite{Paniak, Kupershmidt}
with $2u_2\equiv u, x^2\equiv y,x^3\equiv t$:
\begin{eqnarray}
 \fr{\del u}{\del t}=\frac{1}{4}\fr{\del^3 u}{\del x^3}
+\frac{3}{4}\fr{\del (u\star u)}{\del x}
+\frac{3}{4}\int^x dx^\prime \fr{\del^2 u(x^\prime)}{\del y^2}
-\frac{3}{4}\left[u,\int^x dx^\prime 
\fr{\del u(x^\prime)}{\del y}\right]_\star.
\end{eqnarray}
Important point is that infinite kind of fields $u_3, u_4, u_5,\ldots$
are represented in terms of one kind of field  $2u_2\equiv u$
as is seen in Eq. (\ref{KP_hie}).
This guarantees the existence of NC KP hierarchy
which implies the existence of reductions of
the NC KP hierarchy.
The order of nonlinear terms are determined in this way.

\item NC KdV Hierarchy (2-reduction of the NC KP hierarchy)

Taking the constraint $L^2=B_2=:\del_x^2+u$ for
the NC KP hierarchy, we get the NC KdV hierarchy. 
This time, the following NC Lax hierarchy
\begin{eqnarray}
\label{KdV_hie}
 \frac{\partial u}{\partial x^m}=\left[B_m, L^2\right]_\star,
\end{eqnarray}
include neither positive nor negative power of
(pseudo-)differential operators for the same reason 
as commutative case (See e.g. \cite{Segal}.) and
gives rise to the $m$-th KdV equation for each $m$.
For example, the NC KdV hierarchy (\ref{KdV_hie})
becomes the $(1+1)$-dimensional 
NC KdV equation \cite{DiMH3} for $m=3$ with $x^3\equiv t$
\begin{eqnarray}
\label{ncKdV}
 \dot{u}=\frac{1}{4}u^{\prime\prime\prime}+\frac{3}{4}(u\star u)^\prime,
\end{eqnarray}
and the $(1+1)$-dimensional 5-th NC KdV equation \cite{Toda} 
for $m=5$ with $x^5\equiv t$
\begin{eqnarray}
\dot{u}&=&\frac{1}{16}u^{\prime\prime\prime\prime\prime}
+\frac{5}{16}(u\star u^{\prime\prime\prime}+u^{\prime\prime\prime}\star u)
+\frac{5}{8}(u^{\prime}\star u^{\prime}+u\star u\star u)^\prime,
\end{eqnarray}
where $\dot{u}:=\del u/\del t$.

\item NC Boussinesq Hierarchy (3-reduction of the NC KP hierarchy)

The 3-reduction  $L^3=B_3$ yields the NC Boussinesq hierarchy
which includes the $(1+1)$-dimensional 
NC Boussinesq equation \cite{Toda} 
with $t\equiv x^2$:
\begin{eqnarray}
 \ddot{u}=\frac{1}{3}u^{\prime\prime\prime\prime}
+(u \star u)^{\prime\prime}
+([u,\partial_x^{-1}\dot{u}]_{\star})^\prime,
\end{eqnarray}
where $\ddot{u}:=\del^2 u/\del t^2$ and $\partial_x^{-1}=\int^x dx$.

\item NC coupled KdV Hierarchy (4-reduction of the NC KP hierarchy)

The hierarchy includes 
the $(1+1)$-dimensional NC coupled KdV equation $t\equiv x^3$:
\begin{eqnarray}
  \dot{u}=\frac{1}{4}u^{\prime\prime\prime}
+\frac{3}{4}(u\star u)^\prime
+\frac{3}{4}(\omega-\phi^2)^\prime
-\frac{3}{4}\left[u,\phi^\prime\right]_\star,
\end{eqnarray}
and other two equations with respect to three kind 
of fields $u, \omega,$ and $\phi$, which are 
determined by Eqs. (\ref{KP_hie}) and (\ref{3flow}).
The $x^2$-dependence of the fields is absorbed by 
the fields $\omega, \phi$.

In this way, we can generate 
infinite set of the $l$-reduced NC hierarchies.
If we take other set-up, we can get many other hierarchies.

\item NC Sawada-Kotera hierarchy (3-reduction of the NC BKP hierarchy)

The NC version of BKP hierarchy \cite{JiMi} is obtained
from the NC KP hierarchy 
by the constraint that the constant terms
of $B_m$ for $m=1,3,5,\ldots$ should vanish.
The 3-reduction of the NC BKP hierarchy includes
the $(1+1)$-dimensional 
NC Sawada-Kotera equation with $t\equiv x^5, u\equiv 3u_2$:
\begin{eqnarray}
 \dot{u}+\fr{1}{9}u^{\prime\prime\prime\prime\prime}
+\fr{5}{9}u^{\prime\prime\prime}\star u
+\fr{5}{9}u^{\prime\prime}\star u^\prime 
+\fr{5}{9}u\star u^\prime \star u=0,
\end{eqnarray}
which is new.
\end{itemize}

\noindent
\underline{For $u_1\neq 0$ ($r=1$)}
\noindent
\vspace{3mm}

On commutative spaces, this situation
generates modified KP (mKP) hierarchy and its reductions.
On NC spaces, however, the existence of
them is not always guaranteed.
For the NC KP hierarchy, 
infinite kind of fields are described by one kind of
from the $x^2$-flow equations (\ref{KP_hie}).
However this time the flow equation becomes
\begin{eqnarray}
\label{mKP_hie}
\partial_x^{0})~~~\del_2 u_{1}
&=&u_1^{\prime\prime}+2u_2^{\prime}+2u_1\star u_1^\prime
+2[u_1,u_2]_\star,\nonumber \\
\partial_x^{-1})~~~\del_2 u_{2}
&=&u_2^{\prime\prime}+2u_{3}^{\prime}+2u_1\star u_2^\prime
+2[u_1,u_3]_\star,\nonumber \\
\partial_x^{-2})~~~\del_2 u_{3}&=&\cdots.
\end{eqnarray}
Hence due to the commutator $[u_1,u_k]$,
it is very hard to represent the field $u_k$ 
in terms of $u_1,u_2,\ldots, u_{k-1}$.
The same is true of other flows.
That is why the existence of NC modified KP hierarchy 
is nontrivial. 

Some reduced hierarchies are obtained from constraint conditions.
\begin{itemize}
\item NC mKdV Hierarchy 
(2-reduction of the ``NC mKP hierarchy'')

This time, the 2-reduction constraint $L^2=B_2$
makes it possible to
represent infinite kind of fields $u_2,u_3,\ldots$ 
in terms of one kind of field $2u_1\equiv v$.
The NC mKdV hierarchy
includes the $(1+1)$-dimensional NC mKdV equation for $m=3$
with $t_3\equiv t$:
\begin{eqnarray}
\label{ncmKdV}
\dot{v}
=\frac{1}{4}v^{\prime\prime\prime}
-\frac{3}{8}v\star v^{\prime}\star v+\frac{3}{8}
[v,v^{\prime\prime}]_\star.
\end{eqnarray}

\item NC Burgers Hierarchy \cite{HaTo2}

This is obtained by an irregular reduction.
Putting the constraint $L_{\leq -1}=0$ 
or $L=:\partial_x+v$, the Lax hierarchy (\ref{lax_sato}) 
yields the NC Burgers hierarchy which includes
neither positive nor negative power of differential operator.
For $m=2$, the hierarchy becomes the $(1+1)$-dimensional
NC Burgers equation 
with $t\equiv x^2$:
\begin{eqnarray}
\label{burgers}
 \dot{v}
= [B_2,L]_\star
= [\partial_x^2+2v\partial_x, \partial_x +v]_\star
 =v^{\prime\prime}+2v\star v^{\prime}.
\end{eqnarray}
The NC Burgers equation is linearizable and
easily solved via NC Cole-Hopf 
transformation \cite{HaTo2,MaPa}.
In the linearization, the order of the nonlinear term
play crucial roles. 
This order is automatically realized from Sato's framework.
\end{itemize}

The present discussion is applicable to
the matrix Sato theory where the fields $u_k$ ($k=1,2,\ldots$)
are $N\times N$ matrices.
For $N=2$, the Lax hierarchy
includes the Ablowitz-Kaup-Newell-Segur (AKNS) system \cite{AKNS}, 
the Davey-Stewarson equation, the NLS equation and so on. 
(For commutative discussions, see e.g. \cite{Blaszak}.)

NC version \cite{Toda} of 
the Bogoyavlenskii-Calogero-Schiff (BCS) equation \cite{BCS} 
is also derived from this framework because the Sato theory works well  
on the commutative BCS equation.

\section{Commuting Flows for NC Lax Hierarchies}

First let us show all flows are commuting:
\begin{eqnarray}
\label{commute}
 \del_m\del_n u_k &=& \del_n\del_m u_k
\end{eqnarray}
for any $m,n,k$.
The derivation in this section 
is straightforward as commutative case \cite{Wilson, Segal}
and already discussed in more general situation 
where the products are noncommutative associative products with differentiations.
(See e.g. \cite{Kupershmidt, Wilson2, DrSo}.)

From NC Lax equation (\ref{lax_sato}), we get
\begin{eqnarray}
 \del_m\del_n L =  [\del_m B_n, L]_\star+ [B_n, \del_m L]_\star
=[\del_m B_n, L]_\star+ [B_n, [B_m, L]_\star]_\star.
\end{eqnarray}
Hence 
\begin{eqnarray}
 [\del_m,\del_n] L= [F_{mn},L]_\star,
\end{eqnarray}
where
\begin{eqnarray}
 F_{mn}:=\del_m B_n -\del_n B_m -[B_m, B_n]_\star.
\end{eqnarray}

Now we show the ``zero-curvature equation'' $F_{mn}=0$.
We note that
\begin{eqnarray}
 \del_m B_n&=&\del_m(L^n)_{\geq r}=(\del_m L^n)_{\geq r}\nn
&=&[B_m,L^n]_{\star\geq r}=-[B_m^c,L^n]_{\star\geq r}\nn
&=&-[B_m^c,B_n]_{\star\geq r},
\end{eqnarray}
where the operator $B_m^c$ is the compliment of $B_m$
and defined by 
\begin{eqnarray}
 B_m^{c}:=L^m-B_m,
\end{eqnarray}
and the suffix $r$ is equal to 0 for $u_1=0$ and 1 for $u_1\neq 0$.
Therefore we get
\begin{eqnarray}
\label{nc_ZS}
 F_{mn}&=&-[B_m^c,B_n]_{\star\geq r}+[B_n^c,B_m]_{\star\geq r}
-[B_m, B_n]_\star\nn
&=&-[B_m^c,L^n-B_n^c]_{\star\geq r}+[L^n-B_n,B_m]_{\star\geq r}
-[B_m, B_n]_{\star\geq r}\nn
&=&[B_m^c,B_n^c]_{\star\geq r}\nn
&=&0,
\end{eqnarray}
which implies
\begin{eqnarray}
   \del_m\del_n L &=& \del_n\del_m L.
\end{eqnarray}
Hence Eq. (\ref{commute}) is proved.

We note that the present discussion works well
for arbitrary noncommutativity.
Here we call the equation (\ref{nc_ZS})
the {\it NC Zakharov-Shabat equation}
because it reduces to the usual Zakharov-Shabat equation
in the commutative limit:
\begin{eqnarray}
 \del_m B_n -\del_n B_m -[B_m, B_n]_\star=0.
\end{eqnarray}
Of course, we can get the conjugate of
the NC Zakharov-Shabat equation in terms of $B_n^c$:
\begin{eqnarray}
 \del_m B^c_n -\del_n B^c_m +[B^c_m, B^c_n]_\star=0.
\end{eqnarray}

\section{Conservation Laws for NC Lax Hierarchies}

Here let us prove the conservation laws for
NC Lax equations, which are the main results in the present paper,

First we would like to comment on conservation laws
of NC field equations \cite{HaTo2}.
The discussion is basically the same as commutative case
because both the differentiation and the integration
are the same as commutative ones in the Moyal representation.

Let us suppose the conservation law
\begin{eqnarray}
{\partial \sigma(t,x^i)\over{\partial t}}
=\partial_i J^i(t,x^i),
\end{eqnarray}
where $\sigma(t,x^i)$ and $J^i(t,x^i)$ are called
the {\it conserved density} and the {\it associated flux},
respectively.
The conserved quantity is given by spatial integral 
of the conserved density:
\begin{eqnarray}
Q(t)=\int_{\scr\mbox{space}}d^Dx \sigma(t,x^i),
\end{eqnarray}
where the integral $\int_{\scr\mbox{space}}dx^D$
is taken for spatial coordinates.
The proof is straightforward:
\begin{eqnarray}
{d Q\over{dt}}={\partial
\over{\partial t}}\int_{\scr\mbox{space}} d^Dx \sigma(t,x^i)
=\int_{\scr\mbox{space}} d^Dx \partial_i J_i(t,x^i) 
=\int_{\st{\scr\mbox{spatial}}{\scr\mbox{infinity}}}dS^i J_i(t,x^i) 
=0,
\end{eqnarray}
unless the surface term of the integrand  $J_i(t,x^i)$ vanishes.
The convergence of the integral is also expected because
the star-product naively reduces to the ordinary product 
at spatial infinity due to: $\del_i \sim {\cal{O}}(r^{-1})$ 
where $r:=\vert x \vert$.

For commutative field equations,
the existence of infinite number of conserved quantities 
is expected to lead to infinite-dimensional hidden symmetry 
from Noether's theorem.
For NC field equations, this would be also true and
the existence of infinite number of conserved quantities 
would be special and meaningful, and 
suggest an infinite-dimensional hidden symmetry
deformed from commutative one.

In order to discuss conservation laws for the
NC Lax hierarchies,
let us first calculate the differential of the residue of $L^n$
following G. Wilson's approach \cite{Wilson}: 
\begin{eqnarray}
\label{residue}
 \del_m \res_{-1} L^n = \res_{-1} (\del_m L^n) = \res_{-1} [B_m,L^n]_\star.
\end{eqnarray}
Here we note that 
\begin{eqnarray}
\label{wilson}
 \res_{-1} [f\del_x^p,g\del_x^q]_\star&=&\left(\ba{c}p\\p+q+1\ea\right)
\left(f\star g^{(p+q+1)}-(-1)^{p+q+1}g\star f^{(p+q+1)}\right)\\
&=&\left(\ba{c}p\\p+q+1\ea\right)
\left\{
\left(
\sum_{k=0}^{p+q}
(-1)^k f^{(k)}\star g^{(p+q-k)}\right)^\prime
+(-1)^{p+q}[g,f^{(p+q+1)}]_\star\right\},\nonumber
\end{eqnarray}
where $f^{(N)}:=\del^N f/\del x^N$
Hence we can see that 
on NC spaces, there is an additional term
as a commutator in Eq. (\ref{wilson})
which vanishes in commutative limit.
However as we saw in Sec. 2,
commutators of fields can be represented as
total derivatives, which is very important here.

Let us describe the explicit representations of
the conservation laws. 
From the explicit forms of the Lax pair
\begin{eqnarray}
 L^n&=&\del_x^n+\sum_{l=1}^{\infty}a_{n-l}\del_x^{n-l}\nn
 B_m&=&\del_x^m+\sum_{k=1}^{m}b_{m-k}\del_x^{m-k},
\end{eqnarray}
we can evaluate Eq. (\ref{residue}) as:
\begin{eqnarray*}
\label{gen_conserve}
&&\del_m \res_{-1} L^n
=\res_{-1}[\del_x^m+\sum_{k=1}^{m}b_{m-k}\del_x^{m-k},
\del_x^n+\sum_{l=1}^{\infty}a_{n-l}\del_x^{n-l}]_\star\\
&&=\sum_{l=n+1}^{m+n}\left(\ba{c}m\\l-n-1\ea\right)a_{n-l}^{(m+n-l+1)}
+\sum_{k=1}^{m}\sum_{l=n+1}^{n+1+m-k}
\left(\ba{c}m-k\\l-n-1\ea\right)\nn
&&~~~\times\left\{\left(\sum_{N=0}^{m+n-k-l}
(-1)^{N}b_{m-k}^{(N)}\star a_{n-l}^{(m+n-k-l-N)}\right)^\prime
+(-1)^{m+n-k-l}
\left[a_{n-l},b_{m-k}^{(m+n-k-l+1)}\right]_\star\right\}\nn
&&=\left\{\sum_{l=n+1}^{m+n}\left(\ba{c}m\\l-n-1\ea\right)
a_{n-l}^{(m+n-l)}
+\sum_{k=1}^{m}\sum_{l=n+1}^{\st{n+1+}{m-k}}
\left(\ba{c}m-k\\l-n-1\ea\right)\sum_{N=0}^{\st{m+n}{-k-l}}
(-1)^{N}b_{m-k}^{(N)}\star a_{n-l}^{(m+n-k-l-N)}\right\}^\prime\nn
&&~~~-\sum_{k=1}^{m}\sum_{l=n+1}^{n+1+m-k}
\left(\ba{c}m-k\\l-n-1\ea\right)(-1)^{m+n-k-l}\theta^{ij}\del_i
\left(a_{n-l}\diamond \del_j 
b_{m-k}^{(m+n-k-l+1)}\right)\nonumber
\end{eqnarray*}
This is the generalized conservation laws for
the NC Lax hierarchies.
The RHS contains derivatives in all NC directions.
When we interpreted this as conservation laws,
we have to specify what coordinates correspond to
time and space and introduce the noncommutativities 
in the space-time directions only.

If we identify the coordinate $x^m$ with time $t$,
we get the conserved density as follows:
\begin{eqnarray}
 \sigma=\res_{-1} L^n+\theta^{im}\sum_{k=0}^{m-1}\sum_{l=0}^{k}
(-1)^{k-l}\left(\ba{c}k\\l\ea\right){\mbox{res }}_{-(l+1)} L^n
\diamond \del_i \del_x^{k-l}{\mbox{res }}_{k} L^m,
\label{conservation}
\end{eqnarray}
for $n=1,2,\ldots$, where the suffices $i$ must run
in the space-time directions only.
We can easily see that deformation terms
appear in the second term of Eq. (\ref{conservation})
in the case of space-time noncommutativity.
On the other hand,
in the case of space-space noncommutativity,
the conserved density is given by the residue of $L^n$
as commutative case.

Let us show more explicit representations as follows.

\begin{itemize}

\item In the case that the space-time coordinates 
are $(x,y,t)\equiv (x^1,x^2,x^3)$

The conserved density is given by
\begin{eqnarray}
 \sigma=\res_{-1} L^n+\theta^{im}\sum_{k=0}^{2}\sum_{l=0}^{k}
(-1)^{k-l}\left(\ba{c}k\\l\ea\right){\mbox{res }}_{-(l+1)} L^n
\diamond \del_i \del_x^{k-l}{\mbox{res }}_{k} L^3,
\end{eqnarray}
more explicitly, for $u_1=0$ and $[t,x]=i\theta$, which includes
the NC KP equation with space-time noncommutativity,
the NC KdV equation and so on:
\begin{eqnarray}
\sigma
=\res_{-1} L^n
-3\theta
\left((\res_{-1}L^n)\diamond u_3^\prime
+(\res_{-2}L^n)\diamond u_2^\prime
\right),
\end{eqnarray}
and for $u_1\neq 0$ and $[t,x]=i\theta$, which includes
the NC modified KdV equation and so on:
\begin{eqnarray}
\sigma
&=&\res_{-1} L^n\\
&&+3\theta
\left((\res_{-1}L^n)\diamond (u_2+u_1^2)^{\prime\prime}
-(\res_{-2}L^n)\diamond (u_2-u_1^\prime-u_1^2)^\prime
-(\res_{-3}L^n)\diamond u_1^\prime\right).\nonumber
\end{eqnarray}

\item In the case that the space-time coordinates 
are $(x,t)\equiv (x^1,x^2)$ with $[t,x]=i\theta$

The conserved density is given by
\begin{eqnarray}
 \sigma=\res_{-1} L^n-\theta\sum_{k=0}^{1}\sum_{l=0}^{k}
(-1)^{k-l}\left(\ba{c}k\\l\ea\right){\mbox{res }}_{-(l+1)} L^n
\diamond \del_i \del_x^{k-l}{\mbox{res }}_{k} L^2,
\end{eqnarray}
more explicitly, for $u_1=0$, which includes
the NC Boussinesq equation and so on:
\begin{eqnarray}
\sigma
=\res_{-1} L^n+
2\theta (\res_{-1}L^n)\diamond u_2^\prime,
\end{eqnarray}
and for $u_1\neq 0$:
\begin{eqnarray}
\sigma
=\res_{-1} L^n+
2\theta
\left((\res_{-1}L^n)\diamond u_1^{\prime\prime}
-(\res_{-2}L^n)\diamond u_1^\prime
\right).
\end{eqnarray}
\end{itemize}

We note that for space-space noncommutativity,
conserved quantities (not densities) are
all the same as commutative ones because of Eq. (\ref{integral}).
This is consistent with the present results, of course. 
Furthermore, for $l$-reduced hierarchies,
the conserved densities (\ref{conservation}) 
become trivial for $n=Nl$ ($N=1,2,\ldots$).
The NC Burgers hierarchy is obtained by 
a ``1-reduction'' and contains no negative power
of differential operators.
Hence we cannot generate any conserved density
for the NC Burgers equation in the present approach.
This is considered to suggest that
the NC Burgers equation is not a conservative
system but a dispersive system as commutative case.

We have one comment on conserved densities 
for one-soliton configuration. 
One soliton solutions can always reduce to the commutative ones
because $f(t-x)\star g(t-x)=f(t-x)g(t-x)$ \cite{DiMH3, HaTo2}.
Hence the conserved densities are not deformed
in the NC extension.

The present discussion is applicable to
the NC matrix Sato theory,
including the NC AKNS system, 
the NC Davey-Stewarson equation, 
the NC NLS equation, and the NC BCS equation. 

\section{Conclusion and Discussion}

In the present paper,
we showed that the existence of 
infinite number of conserved densities
for wide class of NC Lax hierarchies
and obtained the explicit representations of them
for both space-space and space-time
noncommutativities.
This suggests that 
NC soliton equations are completely integrable
and infinite-dimensional symmetries 
would be hidden, which would be considered 
as some deformed affine Lie algebras.

In order to reveal what the hidden symmetry is,
we have to first study NC extension of 
Hirota's bilinearization \cite{Hirota}.
This could be realized as a simple generalization of
the Cole-Hope transformation whose extension
to NC spaces are already successful in \cite{HaTo2,MaPa}.
Hirota's bilinearization leads to the theory of tau-functions
which is essential in the discussion of the Lie algebraic
structure of symmetry of the solution space \cite{DJM,JiMi,DJKM, IKT}.
After submission of the present manuscript to this journal,
progress has been reported in e.g. \cite{DiMH5, WaWa2, Sakakibara}.

Our results guarantee that
NC extension of soliton theories would be actually
fruitful and worth studying.
There are many further directions,
such as, the study of relation to $q$-deformations
of integrable systems, NC extension of
the $r$-matrix formalism \cite{Blaszak, FaTa},
the inverse scattering method and
the B\"acklund transformation, and so on.
NC extension of the Ward conjecture \cite{Ward} 
(See also \cite{Conj})
would be also very interesting \cite{HaTo}.
Some NC equations are actually derived from
NC (anti-)self-dual YM equations by reduction
\cite{Legare,HaTo2, NiRa}
and embedded \cite{LPS, LePo, LPS2}
in $N=2$ string theories \cite{OoVa}.
This guarantees that NC soliton equations
would have physical meanings and 
might be helpful to understand new aspects of 
the corresponding string theory.

\subsection*{Acknowledgements}

The author would like to thank 
H.~Awata, 
H.~Ishikawa,
S.~Kakei, M.~Kato, I.~Kishimoto,
A.~Nakamula, S.~Odake,
R.~Sasaki, J.~Shiraishi, K.~Toda, T.~Tsuchida, 
M.~Wadati and S.~Watamura
for useful comments, and 
A.~Dimakis and F.~M\"uller-Hoissen 
for valuable remarks via e-mail correspondence.
He is also grateful to organizers and audiences
during the workshops YITP-W-03-07 on ``QFT 2003'' and
YITP-W-04-03 on ``QFT 2004''
for hospitality and discussion.
This work was supported by 
JSPS Research Fellowships for Young Scientists (\#0310363).

\begin{appendix}

\section{Miscellaneous Formulas}

We present explicit calculations of $L^n$ for $n=1,2,3,4,5$
up to some order of the pseudo-differential operator $\partial_x$.
We can read reduction conditions, e.g. $L^l=B_l$,
and the explicit representations of $\res_{r}L^n$ and $B_m$.

\vspace{2mm}
\noindent
\underline{For $u_1= 0$ ($r=0$):}
\noindent
\begin{eqnarray*}
 L&=& \partial_x + u_2 \partial_x^{-1} 
 + u_3 \partial_x^{-2} + u_4 \partial_x^{-3} 
 + u_5 \partial_x^{-4} + u_6 \partial_x^{-5} 
 + \cdots,\\
 L^2&=&\partial_x^2 + 2u_2 
 + (2u_3+u_2^\prime) \partial_x^{-1} 
 + (2u_4+u_3^\prime+u_2\star u_2) \partial_x^{-2} \nn
 &&+ (2u_5+u_4^\prime+u_2\star u_3+u_3\star u_2-u_2\star u_2^\prime) 
   \partial_x^{-3} \nn
 &&+ (2u_6+u_5^\prime+u_2\star u_4+u_4\star u_2+u_3\star u_3
    -u_2\star u_3^\prime-2u_3\star u_2^\prime+u_2\star u_2^{\prime\prime}) 
   \partial_x^{-4}
 + \cdots,\nn
 L^3&=&\partial_x^3 + 3u_2 \partial_x
 + 3(u_3+u_2^\prime) 
 + (3u_4+3u_3^\prime+u_2^{\prime\prime}+3u_2\star u_2) \partial_x^{-1} \nn
 &&+ (3u_5+3u_4^\prime+u_3^{\prime\prime}
   +3u_2\star u_3+3u_3\star u_2+u_2^\prime\star u_2-u_2\star u_2^\prime) 
   \partial_x^{-2} \nn
 &&+ (3u_6+3u_5^\prime+u_4^{\prime\prime}
    +3u_2\star u_4+3u_4\star u_2+3u_3\star u_3+u_2\star u_2\star u_2\nn
 &&~~~ +u_2^\prime \star u_3-u_2\star u_3^\prime
    +u_3^\prime \star u_2 -4u_3\star u_2^\prime 
    -u_2^\prime\star u_2^\prime+u_2\star u_2^{\prime\prime}) 
   \partial_x^{-3}
 + \cdots,\nn
 L^4&=&\partial_x^4 + 4u_2 \partial_x^2
 + (4u_3+6u_2^\prime) \partial_x
 + (4u_4+6u_3^\prime+4u_2^{\prime\prime}+6u_2\star u_2) \nn
 &&+ (4u_5+6u_4^\prime+4u_3^{\prime\prime}+u_2^{\prime\prime\prime}
   +6u_2\star u_3+6u_3\star u_2+4u_2^\prime\star u_2+2u_2\star u_2^\prime) 
   \partial_x^{-1} \nn
 &&+ (4u_6+6u_5^\prime+4u_4^{\prime\prime}+u_3^{\prime\prime\prime}
    +6u_2\star u_4+6u_4\star u_2+6u_3\star u_3
    +4u_2\star u_2 \star u_2 \nn
 &&~~~ +4u_2^\prime \star u_3+2u_2\star u_3^\prime
    +4u_3^\prime \star u_2-4u_3\star u_2^\prime
    -u_2^\prime\star u_2^\prime+u_2^{\prime\prime}\star u_2
    +u_2\star u_2^{\prime\prime}) 
   \partial_x^{-2}
 + \cdots,\nn
 L^5&=&\partial_x^5 + 5u_2 \partial_x^3
 + 5(u_3+2u_2^\prime) \partial_x^2
 + 5(u_4+2u_3^\prime+2u_2^{\prime\prime}+2u_2\star u_2)\partial_x \nn
 &&+ 5(u_5+2u_4^\prime+2u_3^{\prime\prime}+u_2^{\prime\prime\prime}
   +2u_2\star u_3+2u_3\star u_2+2u_2^\prime\star u_2+2u_2\star u_2^\prime) 
   \nn
 &&+ (5u_6+10u_5^\prime+10u_4^{\prime\prime}+5u_3^{\prime\prime\prime}
    +u_2^{\prime\prime\prime\prime}
    +10u_2\star u_4+10u_4\star u_2+10u_3\star u_3+10u_2\star u_2\star u_2
     \nn
 &&~~~ +10u_2^\prime \star u_3
    +10u_2\star u_3^\prime+10u_3^\prime \star u_2 
    +5u_2^\prime\star u_2^\prime+5u_2^{\prime\prime}\star u_2
    +5u_2\star u_2^{\prime\prime}) 
   \partial_x^{-1}
 + \cdots.\nonumber
\end{eqnarray*}
\noindent
\underline{For $u_1\neq 0$ ($r=1$):}
\noindent
\begin{eqnarray*}
 L&=& \partial_x + u_1 +u_2 \partial_x^{-1} 
 + u_3 \partial_x^{-2} + u_4 \partial_x^{-3} 
 + u_5 \partial_x^{-4} + u_6 \partial_x^{-5} 
 + \cdots,\\
 L^2&=&\partial_x^2 + 2u_1\partial_x
 + (2u_2+u_1^\prime+u_1^2)
 + (2u_3+u_2^\prime+u_1\star u_2+u_2\star u_1) \partial_x^{-1}\\ 
 &&+ (2u_4+u_3^\prime+u_1\star u_3+u_3\star u_1+u_2\star u_2
     -u_2\star u_1^\prime) \partial_x^{-2}\\
 &&+ (2u_5+u_4^\prime+u_1\star u_4+u_4\star u_1
   +u_2\star u_3+u_3\star u_2-2u_3\star u_1^\prime-u_2\star u_2^\prime
   +u_2\star u_1^{\prime\prime}) 
   \partial_x^{-3}\\
&&+ \cdots,\\
 L^3&=&\partial_x^3 + 3u_1\partial_x^2
   + 3(u_2+u_1^\prime+u_1\star u_1)\partial_x \\
&& + (3u_3+3u_2^\prime+3u_1^{\prime\prime}+3u_1\star u_2+3u_2\star u_1
    +u_1^\prime\star u_1+2u_1\star u_1^\prime +u_1\star u_1\star u_1) \\
&& + (3u_4+3u_3^\prime+u_2^{\prime\prime}+3u_1\star u_3+3u_3\star u_1
 +3u_2\star u_2+u_1^\prime\star u_2
 +2u_1\star u_2^\prime\\
&&~~~  +u_2^\prime \star u_1 -2u_2\star u_1^\prime
  +u_1\star u_1\star u_2+u_1\star u_2\star u_1
  +u_2\star u_1\star u_1) \partial_x^{-1}  + \cdots,\\
 L^4&=&\partial_x^4 + 4u_1\partial_x^3 
+ (4u_2+6u_1^\prime +6u_1\star u_1) \partial_x^2\\
&& + (4u_3+6u_2^\prime+4u_1^{\prime\prime}+6u_1\star u_2+6u_2\star u_1
    +4u_1^\prime\star u_1+8u_1\star u_1^\prime+4u_1\star u_1\star u_1) 
   \partial_x\\
&&+ (4u_4+6u_3^\prime+4u_2^{\prime\prime}+u_1^{\prime\prime\prime}
    +6u_1\star u_3+6u_3\star u_1+6u_2\star u_2\\
&&~~~    +4u_1^\prime\star u_2+6u_1\star u_2^\prime+4u_2^\prime \star u_1
  -2u_2\star u_1^\prime+2u_1^{\prime\prime}\star u_1
  +2u_1\star u_1^{\prime\prime}+3u_1^\prime\star u_1^\prime\\
&&~~~ +4u_1\star u_1\star u_2+4u_1\star u_2\star u_1+4u_2\star u_1\star u_1
  +u_1^\prime \star u_1\star u_1 +2u_1\star u_1^\prime\star u_1\\
&&~~~   +3u_1\star u_1 \star u_1^\prime +u_1\star u_1\star u_1\star u_1)
 + \cdots,\\
 L^5&=&\partial_x^5 +5u_1\partial_x^4+ 5(u_2+2u_1^\prime+2u_1\star u_1) 
   \partial_x^3\\
&& + 5(u_3+2u_2^\prime+2u_1^{\prime\prime}+2u_1\star u_2+2u_2\star u_1
    +2u_1^\prime\star u_1+4u_1\star u_1^\prime+2u_1\star u_1\star u_1) 
    \partial_x^2\\
&& + (5u_4+10u_3^\prime+10u_2^{\prime\prime}+5u_1^{\prime\prime\prime}
    +10u_1\star u_3+10u_3\star u_1+10u_2\star u_2\\
&&~~~    +10u_1^\prime\star u_2+20u_1\star u_2^\prime
  +10u_2^\prime \star u_1+4u_2 \star u_1^\prime
  +6u_1^{\prime\prime}\star u_1 +15u_1^\prime\star u_1^\prime
  +11u_1\star u_1^{\prime\prime}
  \\
&&~~~  +10u_1\star u_1\star u_2+10u_1\star u_2\star u_1
    +10u_2\star u_1\star u_1\\
&&~~~     +5u_1^\prime \star u_1\star u_1 +10u_1\star u_1^\prime\star u_1
    +15u_1\star u_1 \star u_1^\prime +5u_1\star u_1\star u_1\star u_1)
\partial_x 
 + \cdots.
\end{eqnarray*}

\end{appendix}

\end{document}